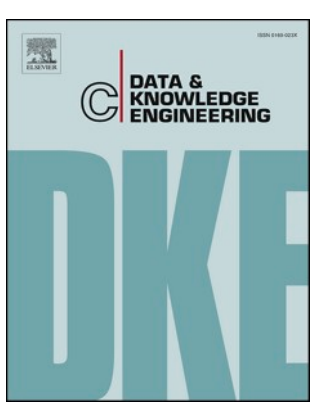

# A realist theory of digital objects, digital systems, and digitalized systems

Roman Lukyanenko [a,*], Ron Weber [b,c]

[a] University of Virginia, Charlottesville, VA 22903, United States
[b] Monash University, Faculty of Information Technology, 25 Exhibition Walk (Building 63), Clayton Victoria 3800, Australia
[c] The University of Queensland, UQ Business School, Level 2, Colin Clark Building, St Lucia Queensland 4072, Australia

A B S T R A C T

As human reliance on information technology (IT) increases, having a clear, precise, and comprehensive understanding of the nature of digital objects, digital systems, and digitalized systems becomes more critical. Otherwise, our ability to research, manage, control, and make reliable predictions about them will be limited. Accordingly, in this paper, we propose a new theory of digital objects, digital systems, and digitalized systems based on the adoption, adaptation, and extension of existing theories of ontology, semantics, and semiotics. The theory provides precise explanations of the nature of digital objects, digital systems, and digitalized systems. Ours is a realist theory that does not countenance the independent existence of nonmaterial or hybrid objects in the world. Accordingly, we are at odds with much of the prevailing discourse about digital phenomena. We show how our theory generates different insights and predictions from this the dominant discourse on the nature of the digital. Our predictions lay the groundwork for further empirical studies on the design, use, and impact of IT on individuals, organizations, and society.

Although we jump to notice the role of power, identity, and other familiar social constructs in light of new technologies, such as large language models in organizations, *few bother to delve into the technology deeply enough to understand precisely how the new technology is different from previous* as it interacts with social variables. ([1], p. 1518) (our emphasis)

## 1. Introduction

As digitalized products and services have become more pervasive and complex, scholars have attempted to understand their nature (e.g., [2–5]). In part, their motivation has been to enable better understanding of the ways that digitalization has diffused and infused

* Corresponding author.
*E-mail addresses:* romanl@virginia.edu (R. Lukyanenko), Ron.Weber@monash.edu, uqrweber@uq.edu.au (R. Weber).

and its growing impact on individuals, organizations, and society (e.g., [1,6–14]).

Despite these efforts, many issues pertaining to the nature of digitalization remain unclear and contested. For example, in information systems (IS) research, digital objects are often deemed to be *hybrid* objects that are partially composed of material and nonmaterial objects (e.g., [4]). As such, do we now have a new type of reality—a hybrid reality— that is the "digital world"? If so, how do the three worlds relate—the material, the nonmaterial, and the digital? Digital objects have also been characterized as having a "dubious ontology" ([15], p. 364) and existing in the world as "quasi-objects" that lack a "clear identity" and are "performative" [3]. How, then, do objects characterized in these ways interact with other objects?

Answers to such questions are widely consequential. As human reliance on information technology (IT) increases, having a clear, precise, and comprehensive understanding of the nature of digital objects, digital systems, and digitalized systems becomes more critical. Otherwise, our ability to research, manage, control, and make reliable predictions about them will be limited.

Accordingly, in this paper, we propose a new theory of digital objects. We extend this theory to also cover digital systems and digitalized systems. We define these terms carefully. Furthermore, we provide precise explanations of the nature of digital objects, digital systems, and digitalized systems. We share Yoo et al.'s concern expressed in our opening quote—namely, the need for more scholars to delve deeply into the technology and understand how it differs from prior technologies.

We have two goals for our theory: (a) to enable better theorizing about and empirical research on digital objects, digital systems, and digitalized systems, and (b) to provide a foundation for their better design, management, use, and maintenance. Absent a clear, precise, and comprehensive theory, the nature of and strengths and weaknesses of digital objects, digital systems, and digitalized systems and how they relate to other objects (human and non-human) cannot be understood fully and will remain contested.

Importantly, we articulate a *realist, materialist* theory of digital objects, digital systems, and digitalized systems. This contrasts with much of the prevailing discourse (increasingly the hegemony) that conceives of digital objects, digital systems, and digitalized systems, at least in part, using a nonmaterial (or immaterial) lens (e.g., [1,2,4,16–21]). Instead, we seek to show that theories about digital objects, digital systems, and digitalized systems that are rooted in nonmaterial worlds, even if only partially, are likely to be problematic.

Our paper proceeds as follows. First, we discuss prior work to better understand current views about the nature of digital objects, digital systems, and digitalized systems. Second, we discuss the fundamental nature of our theory. Third, we describe the major components of our theory. Fourth, we present some illustrative implications of our theory for research and practice. Finally, we present some conclusions.

## 2. Prior conceptualizations of digital objects, digital systems, and digitalized systems

Before undertaking our work, we conducted a broad survey of previous attempts to understand the nature of digital objects, digital systems, and digitalized systems. Appendix A provides an overview of some key literature that we examined. Below, are our main conclusions about this literature.

First, many authors skirt the issue of defining digital objects, digital systems, and digitalized systems explicitly and precisely. Their focus instead is on their putative properties, such as "low cost," "modular," and "reprogrammable" (e.g., [5,7,22]). In particular, a common focus in the literature, often unwittingly, is the *mutual* (or relational) properties of digital objects, digital systems, and digitalized systems—those properties that arise because they interact with other objects (e.g., [23–25]). For instance, "editability" is often listed as a property of digital objects, but it is not a property of read-only integrated circuits (chips). Instead, "editability" is a mutual property of a digital object that can be changed, a tool that can be used to enact changes to a digital object, and an agent that can use the tool properly. Describing the properties of digital objects based upon their mutual properties complicates the quest to distil their fundamental nature, however, because digital objects, digital systems, and digitalized systems interact with many different types of objects. Thus, they may have different and changing mutual properties depending on the objects with which they interact.

Second, digital objects, digital systems, and digitalized systems are covered by different terms—for instance, digital artifacts, digital technologies, technological objects, digital systems, informational objects, quasi-objects, bitstrings, and new media objects (e.g., [1,4,15,26–28]). These diverse terms leave much room for confusion, because it is unclear how they relate and whether they have the same or different meanings.

Third, much of the literature vacillates between whether digital objects, digital systems, and digitalized systems have material or nonmaterial components (e.g., [2,3,27,29–32]). For instance, Faulkner and Runde [4] propose an ontology that has three types of digital objects in the world: (a) material (e.g., smartphones and servers); (b) nonmaterial (e.g., syntactic objects such as newspaper articles and software); and (c) hybrid (objects that have both material and nonmaterial components, such as a laptop computer with software). The nature of materiality and nonmateriality and these different types of objects is unclear. Moreover, Mahner and Bunge [[33], p. 110] note that "nobody has ever come up with a satisfactory conception of how material and immaterial objects may conceivably interact."

Fourth, some authors focus on function and form and background the underlying material substrate (the level structure of basic and composite material components) [2,32]. They argue that the evolution of digital objects, digital systems, and digitalized systems has led to "a growing emancipation" of their form and function "from the materials in which they are entangled" ([34], p. 69). While analytically this argument is seductive, contrariwise, we argue below that it is problematic because the properties of digital objects, digital systems, and digitalized systems (e.g., their form and function) are tied inextricably to their material substrates.

## 3. Nature of our theory

Our theory differs from the work done in prior literature on digital objects, digital systems, and digitalized systems in four important ways. First, we ground our theory in fundamental, primitive constructs. Second, we define our constructs *formally* so their meanings are clear. Third, we show how our constructs are related such that our theory constitutes a coherent whole. Fourth, we define digital objects, digital systems, and digitalized systems based only on those *intrinsic* properties that we contend are common to all of them (whatever their type) and reflect their inherent nature.

The theory that we propose is a Type-I "theory for analyzing"—one that seeks to account for "what is" and not explanation, prediction, or design and action ([35], p. 622). Type-I theories are employed to establish a general foundation for future studies and increase conceptual clarity and precision. As a theory for analyzing, our theory does not contain "causal relationships" ([35], p. 620). Nonetheless, it has well-defined, interrelated constructs that can be used to explain and predict many types of phenomena associated with digital objects, digital systems, and digitalized systems. We provide some examples of how these outcomes can be achieved in Section 5 below.

To ensure rigor and consistency, we have grounded our theory in the general ontology, semantics, and semiotics developed by the philosopher-physicist, Mario Bunge [36–40]. Bunge's is one among several prominent ontologies that elucidate the nature of reality (e.g., [41–44]). We chose Bunge's ontology because his works provide explicit formalization of the relationships between ontological, semantic, and semiotic constructs, which are the important relationships that underpin our theory. Furthermore, Bunge's work has been used extensively in prior IS research (e.g., [45]). In our view, it has enabled novel, useful outcomes to be produced. For example, it has been used to predict the strengths and weaknesses of several conceptual modeling grammars, and some of these predictions have been supported empirically (e.g., [46]).

More generally, we seek to enact the epistemological doctrine of *realism*—namely, finding the best ways to *represent* reality that "is composed exclusively of material things" ([47], p. ix). As realists, we believe "the world can be known," but nonetheless "all knowledge of facts is incomplete and fallible, and much of it is indirect" ([48], p. 29).

## 4. Components of our theory

Our theory has five components. The first component establishes some basic constructs we use in our theory. The second component frames our conception of digital objects, digital systems, and digital components of digitalized systems as types of signs. Equipped with these two foundational components, the third component formalizes and explains the nature of digital objects, digital systems, and digitalized systems. The fourth component examines how digital objects, digital systems, and the digital components of digitalized systems are interpreted through the "lens" of symbols. The fifth component examines the nature of properties, functions, forms, and mechanisms in digital objects, digital systems, and digitalized systems.

### *4.1. Component I: ontological foundations*

To understand the nature of digital objects, digital systems, and digitalized systems, we first draw upon Bunge's [38,39] ontology—a theory about the elementary constituents of reality. His ontology is a component of his philosophy of *scientific materialism.* It is (a) *scientific*, "because it draws its inspiration from science and is tested as well as modified by the advancement of science" ([47], p. xiv); and (b) *materialist*, because it is based on "the thesis that everything that exists is material—or, stated negatively, that immaterial objects such as ideas have no existence independent of material things such as brains" ([47], p. 17).[1]

The basic unit in Bunge's ontology is an *object*, which he describes as follows ([49], p. 199): An object is "[w]hatever can exist, be thought about, talked about, or acted upon. The most basic, abstract, and general of all philosophical concepts, hence undefinable. … Objects can be individuals or collections, concrete (material) or abstract (ideal), natural or artificial."[2]

Bunge [[38], p. 117] partitions "objects" into two classes: *things* (or *material objects*) and *constructs* (or *immaterial objects*).[3] This distinction is strictly *methodological*, a convenient abstraction in the world made solely of material objects.

***Definition 1: Material Object (*[47]*, p. 22):***

**"An object $x$ is a *material object* (or *entity*) if, and only if, for every reference frame $y$, if $S_y(x)$ is a state space for $x$, then $S_y(x)$ contains at least two elements. Otherwise $x$ is an *immaterial object* (or *nonentity*)."**

Thus, the defining property of *material objects* (things) is that inherently they are capable of changing their state. They can change their state because all things possess the property of having energy.[4] Indeed, Bunge [[48], pp. 26–27] observes that "since being changeable is the same as possessing energy, the predicate 'is material' turns out to be coextensive…with 'has energy'". The properties of things are also material, but they are material *derivatively* because of the things they characterize ([48], p. 11). Thus, they are *real*, but not "autonomously real" ([38], p. 99).

[1] See Cordero [82,67] for a summary of and commentary on Bunge's scientific realism.

[2] We note the generality of the term *object* as we proceed to understand the nature of digital *objects*.

[3] This distinction between things and constructs lies at the heart of Bunge's ontology and informs our analysis below of the nature of digital objects, digital systems, and digitalized systems.

[4] Bunge [[83], p. 459] postulates that energy is the only universal property of matter. He defines energy "as the property of being changeable in some respect" (p. 461).

*Constructs* are objects that exist in the brain of a thing (e.g., the number "3", an imagined wizard, or a conceptual model used to inform the design of a database). They are *emergent properties* of the brain—properties that arise as a result of a brain's components (e.g., its chemical elements and neurons) interacting in particular ways. These interactions and the changes that occur (brain processes) ultimately manifest as thoughts [50]. In this regard, Bunge [[47], p. 168] argues that "*every construct is an equivalence class of thoughts* (brain processes of a certain type)."[5] Nonetheless, as a materialist, Bunge [[47], p. 168] underscores that "*our construal of constructs is materialist* because it is rooted in the notion of thoughts as brain processes" (our emphasis).

For *methodological* purposes, however, Bunge conceptualizes constructs as *immaterial* because constructs by themselves cannot change their state ([47], p. 163; [48], p. 10). By methodologically separating things and constructs, Bunge is seeking to gain communicative and cognitive efficiency. Humans can then talk about and think about ideas, theories, predicates, and other constructs as if they exist on their own, without each time having to specify their status as emergent properties of a brain and explain the process of emergence. Considered independently of their brain substrate, constructs can also be modelled as having their own properties—*conceptual* (*formal*) properties[6] such as logical consistency and abstract structure.

Consistent with Bunge's ontology, in our theory of digital objects, digital systems, and digitalized systems, we retain the thing-construct distinction, but only as a convenient stratagem. Moreover, we underscore that *our usage of construct as methodologically dissimilar from material things differs from extant theories that consider constructs as having an independent, immaterial existence.*

Table 1 provides some key constructs from Bunge's ontology, and some extensions of Bunge's ontology (e.g., [51]), that underpin our theory. We do not labor our discussion of them because, as we indicated above, they have already been extensively canvassed in the IS literature. Good summaries of some of the components are also available elsewhere (e.g., [52,53]).

### 4.2. Component II: semantics and semiotics

Equipped with some of Bunge's basic ontological constructs, we now adopt and adapt elements of his theories of semantics and semiotics (e.g., [36,37,40]) to develop the following constructs: *sign, symbol,* and *translator.*

*Definition 2: Sign:*

A *sign* is a thing that possesses a particular property (or properties) that maps to one or more other properties of the same thing, another object (thing or construct), or the properties of another object.

Based on our definition of "sign," we note the following:

- Signs are *material* objects—a *thing* possessing a particular material (real) property (or properties).
- The properties of a *thing* include its states, events, transformations, processes, laws, and history of its states and events. Any of these properties might mean that a thing is considered a *sign.*
- The object to which a sign maps, either directly or indirectly, may be a construct, a thing, or a property of a construct or thing. We discuss the nature of these mappings in Component IV of our theory below.
- Signs are often classified as "natural" (e.g., dark storm clouds) or "artificial" (e.g., a stop sign) (Fig. 1) ([54], pp. 86–87, 113). As digital objects are human artifacts, our focus is artificial signs.
- Artificial signs are *created* according to rules that constrain how the sign is mapped to its associated object. Rules also constrain how an artificial sign is to be properly *interpreted* ([37], pp. 1–3). The rules may vary in explicitness and formality.

Artificial signs can be classified as "iconic" or "non-iconic" (Fig. 1). Non-iconic signs are called "symbols" and are interpreted by convention rather than by similarity to a referent. This classification is observer dependent. For instance, a stop sign might be an iconic sign for a human but a non-iconic sign for an autonomous vehicle. In our theory, we focus on *symbols* because modern digital technologies rely on non-iconic signs (e.g., binary digits).

*Definition 3: Symbol:*

A *symbol* is a thing that is an artificial, non-iconic sign (Fig. 1).

A symbol often belongs to a *semiotic system* that enables it to be concatenated with other symbols (via a grammar) to produce new symbols that map to new, sometimes more complex, objects and properties of objects [40].

Using a symbol directly is often impractical. For example, we may need to map one symbol to another for more durable or convenient storage (e.g., digitize a paper book to store it on a disk or convert a document from MS Word to Adobe PDF format). This mapping requires a *translator.*

*Definition 4: Translator:*

A *translator* is a *thing* that maps (according to a set of rules) one object to another object.[7]

Examples of translators are (a) humans or machines that take text in one language and translate it into text in another language, (b) Zip software that compresses files, (c) sensors that read environmental conditions and produce bitstring values, and (d) codec software, which is part of an operating system, that converts text, numbers, and other content (e.g., video) to bitstring values (so they can be

[5] A construct is an equivalence class of brain processes because different processes in different brains can produce the same construct (e.g., the concept of an integer number or a particular property that characterizes some thing).

[6] Conceptual or formal properties are independent of facts about the real world ([49], p. 110).

[7] Our definition of "translator" is based upon Bunge's [[36], p. 97] definition of a "transformation formula": "DEFINITION 3.5: A statement in a theory *T* is called a *transformation formula* of *T* iff the statement relates equivalent representations of the same factual items."

**Table 1**
Summary of Some Foundational Ontological Constructs (primarily from Bunge [38]) that Underpin Our Theory.

*Objects* are the fundamental, undefined ontological unit. They are partitioned into *things* (material objects) and *constructs* (immaterial objects). This distinction is for methodological convenience because, in reality, constructs are material too (they represent a material property of a material brain).

*Things* are characterized by *material (real) properties* (e.g., their shape in spacetime). *Constructs* are characterized by *formal properties* (e.g., their logical consistency).

Two types of properties possessed by a *thing* are (a) *intrinsic properties*, which are possessed by the thing alone (e.g., its rest mass), and (b) *mutual properties*, which are possessed by a thing only in relation to another thing (e.g., the number of children begot by a male with a particular woman).

A *property in general* is a property possessed by a set of things (e.g., the property age possessed by humans). A *property of a particular* is a property possessed by a particular thing (e.g., the human "John" has an age of 42 years).

Unless they are deemed to be atomic, *things* and *constructs* can be conceptualized as *systems*—objects with interconnected internal components. Systems have a *composition*:

- composition of a system thing: other things
- composition of a system construct: other constructs.

The composition of systems exists in the form of a *level structure*, which shows how some components are composed of other components.

A *state* of a thing are the properties of a thing at some point in a reference frame (e.g., a spacetime coordinate). A state can be modelled as the sequence of the values of properties at some point in a reference frame.

A *transformation* is a mapping from a domain comprising states to a codomain comprising states. For example, plugging a laptop, transforms its battery from a low battery charge (the domain, the starting state) to a fully charged battery (the codomain, the end state).

An *event* in a thing is a change of state of the thing enacted by a particular transformation.

A *process* in a thing is a sequence of events enacted by a particular transformation.

A *history* of a thing is its sequence of states in spacetime.

An *internal process* of a thing is one that does not result in a state change in another thing that is in the environment of the thing.

An *external process* of a thing is one that results in a state change in another thing that is in the environment of the thing.

A *law* restricts the states that a thing can assume or the events it can undergo.

The totality of properties of a thing includes its states (properties in all reference frames), events, processes, transformations, laws, and the history of its states and events. The set of all properties of a thing constitutes its *form*.

One thing *interacts* with another thing if its history is not independent of the *history* of the other thing. One thing is *independent* of another thing if its history is not related to the history of the other thing.

Things that are systems have interacting components that are things. Some of these components interact with things in the environment of the system. Only things can interact with one another.

Things that are systems have *hereditary properties* (properties also possessed by at least one of their components) and *emergent properties* (properties possessed only by the system as a whole).

Constructs that are systems also have hereditary and emergent properties. For instance, a theory may be logically true or false in the same way that each of its propositions may be logically true or false (hereditary property). A theory may also possess the emergent property of "parsimony."

A system's *function: what* a system does (its purpose) via its internal processes to achieve its purpose (described in terms of a desired state of the system).

A system's *mechanisms: how* a system enables its function to be achieved (the processes it uses).

Systems composed of things have four types of *complexity* ([49], p. 47):

- *compositional* (number and types of components)
- *environmental* (number, types, and intensities of external couplings)
- *structural* (number, types, and intensities of internal couplings)
- *mechanismic* (number and types of processes that make a system work—whatever constrains or enables change in the system).

processed and stored on a hard drive as magnetic flux).

### 4.3. Component III: digital phenomena

We now synthesise the ontological foundations and semantics and semiotics components of our theory, along with their corresponding constructs, to uncover the fundamental nature of digital objects, digital systems, and digitalized systems. We formalize the following constructs: *bitstring construct, instance of bitstring construct, bitstring symbol, digital object, digital system*, and *digitalized system*. Fig. 2 shows how some key constructs in our theory are related.

*Definition 5: Bitstring Construct*:

A *bitstring construct* is a mathematical *sequence* of zeros and ones (binary digits).

Based on our definition of "bitstring construct," we note the following:

- Bitstring constructs are a type of *digital construct* because they use only a proper subset of digits (0–1 of 0–9).
- When conditioned by the probability of obtaining a given outcome (0 or 1), bitstring probability vectors are known as Hilbert spaces, which are the basis for quantum computing ([55], pp. 205–206).
- As a *construct*, bitstring constructs have *formal* properties (e.g., a *hamming weight*—the number of non-0 symbols (i.e., 1 s for bitstrings), which is a property that is important for error-correction algorithms).

*Definition 6: Instance of a Bitstring Construct*:

An instance of a *Bitstring Construct* is a specific sequence of zeros and ones.

For instance, in the ASCII code, the Latin letter "A" is represented in bits as "01000001," which is a *specific* sequence of zeros and ones (an instance of a bitstring construct).

Based on our definition of "Instance of a Bitstring Construct," we note the following:

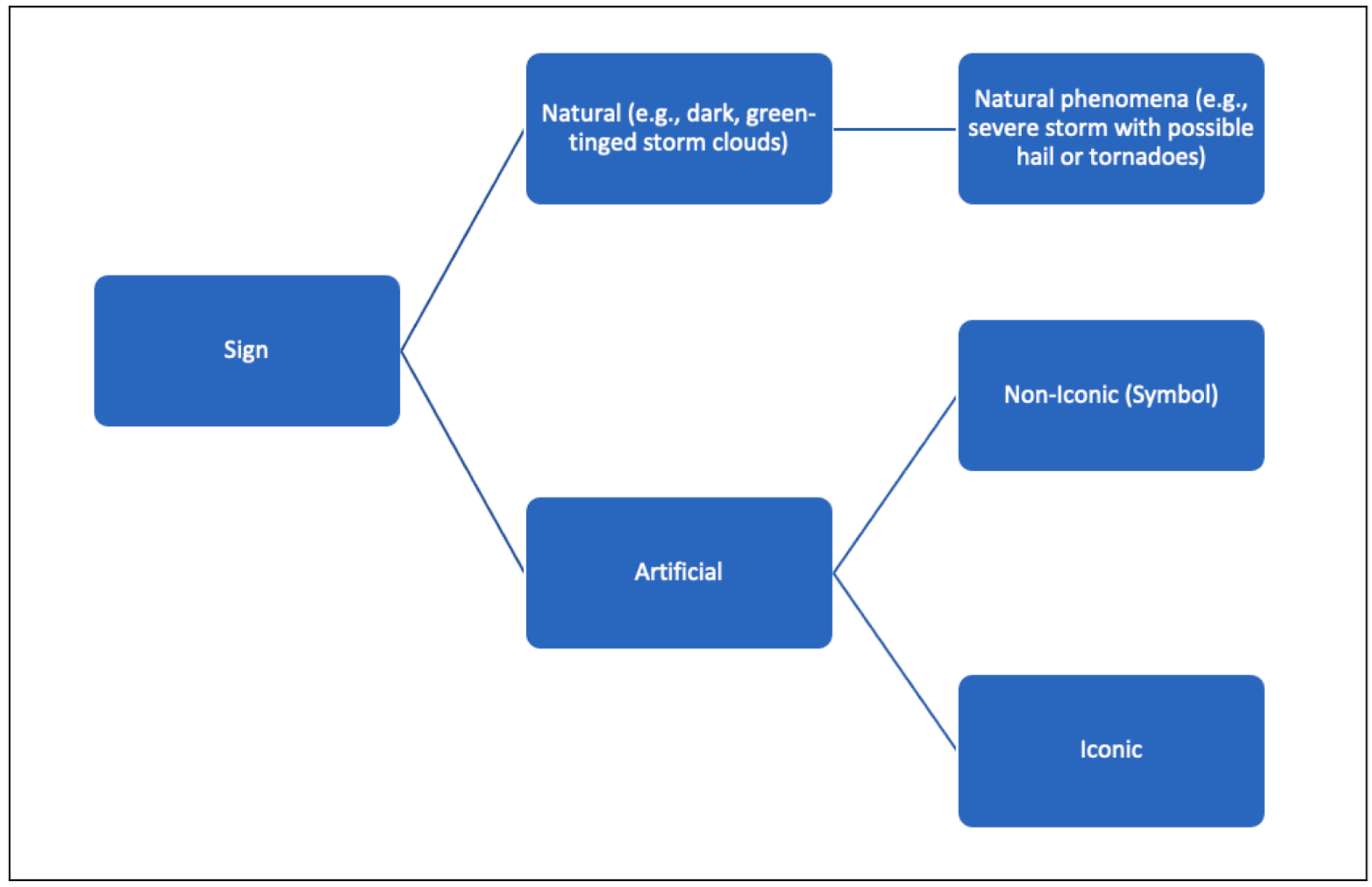

**Fig. 1.** Types of Signs (based on [40], pp. 341–343).

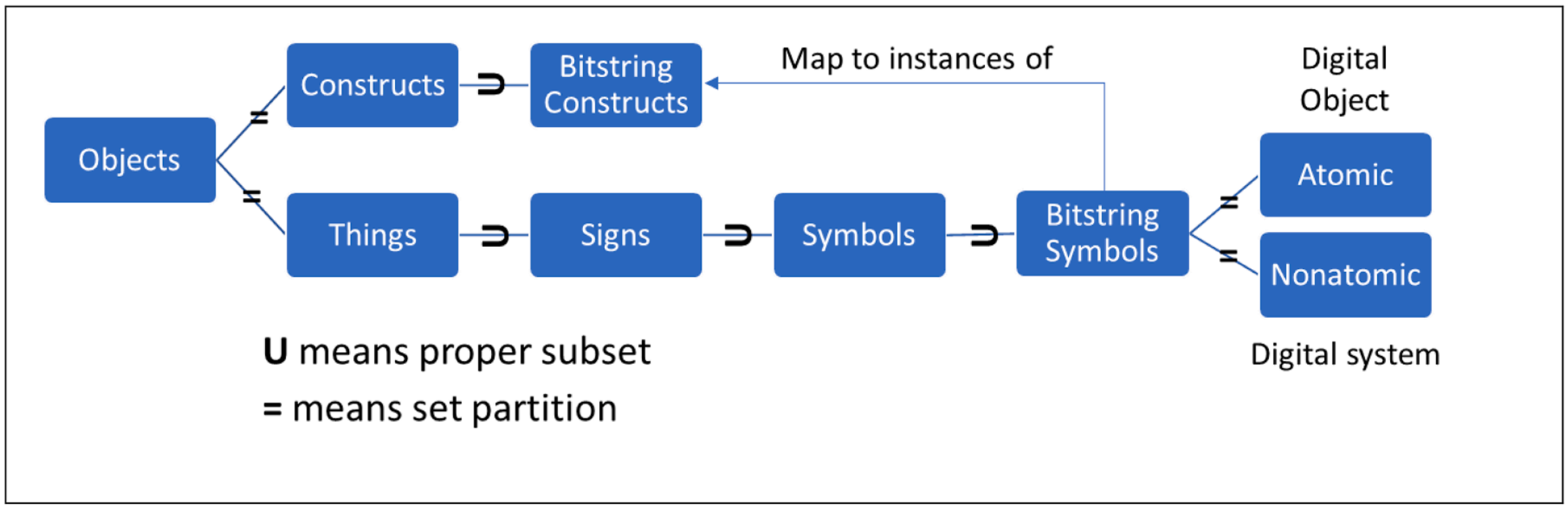

**Fig. 2.** Relationships among some key digital constructs.

- As an instance of a mathematical sequence, an instance of a bitstring construct (in essence, a bitstring value) may have zero, one, or many elements. Also, their elements (ones and zeros) can be repeated.
- To interpret what an instance of a bitstring construct represents, a *translator* must be used to *parse* it. Similarly, to convert an object (thing or construct) to an instance of a bitstring construct, a translator must be used.

*Definition 7: Bitstring Symbol:*

A *bitstring symbol* is a thing that has at least one intrinsic, hereditary, or emergent property that maps the symbol to an instance of a bitstring construct.

For instance, a disk, which is a "thing," has surface properties allowing magnetic fluxes that map to the bits "01000001," which as an ASCII code stands for the Latin letter "A". Thus, a disk is a bitstring symbol.

Based on our definition of "bitstring symbol," we note the following:

- Bitstring symbols are *digital symbols* because they map to digits, although to the restricted domain of zeros and ones (they are a proper subset of digital symbols).
- Working with bitstring symbols directly is usually impractical because their properties (e.g., individual storage locations) have values that are imperceptible to the human eye (e.g., magnetic charges, electric charges, electron spin, or photon polarization). As a result, bitstring symbols often require additional components that enable their property values to be accessed (e.g., reading heads on a hard drive and corresponding circuitry).
- As a *thing*, bitstring symbols have *real* properties (e.g., leakage of the charge representing a bit in dynamic random-access memory (DRAM)) (Kim et al., [56]).

What, therefore, is the meaning of the terms "digital object," "digital system," and "digitalized system"? Based on our analysis of bitstring constructs and bitstring symbols, our theory reveals the problematic nature of these terms.

In particular, recall that our theory distinguishes between two types of "objects": *constructs* (ideas in a mind, emergent property of brains) and *things* (objects that exist independently of a mind). Thus, the term "digital object" is ambiguous because it could mean either a "*digital construct*" (e.g., mathematical elements that are only digits) or a "*digital thing*" (e.g., an electronic display that shows where all horses are placed at the end of a race). Moreover, the term "digital object" implies the conceptual or real property that characterizes the object maps to digits (0–9) and not bits (0,1). In short, "digital objects" are *not* "bitstring objects" (objects with properties that map only to bits). Thus, they are misnamed.

To disambiguate these terms, we define them as follows:

*Definition 8a: Digital Material Object (DMO):*

A *digital material object* is an atomic digital symbol.

A digital material object exists because, for some types of modelling purposes, it does not need to be decomposed into its components. For instance, the internal memory of a laptop computer is composed of transistors. Nonetheless, statements about the size of the laptop's internal memory in gigabytes or mean time between failure can be made by considering its internal memory as an atomic digital material object.

*Definition 8b: Digital Conceptual Object (DCO):*

A *digital conceptual object* is an atomic digital construct.

A digital conceptual object exists because, for some types of modelling purposes, it does not need to be decomposed into its components. For instance, conceptually, a bitstring message to be communicated over a network can have several components such as the sender's identity, the receiver's identity, the component containing the semantics to be communicated to the final receiver, an error detection code, and a public cryptographic key. Nonetheless, statements about the message's overall length in bits do not require the message to be decomposed into its components.

*Definition 9a: Digital Material System (DMS):*

A *digital material system* is an artificial material system whose components are only digital material objects or other digital material systems.

For example, a query processor, storage engine, scheduler, and query optimizer, which are running on a particular server, constitute components (subsystems) of a digital material system—namely, a database management system (DBMS).

*Definition 9b: Digital Conceptual System (DCS):*

A *digital conceptual system* is an artificial conceptual system whose components are only digital conceptual objects or other digital conceptual systems.

For example, a bitstring public key (conceptual object), a bitstring private key (conceptual object), and a bitstring algorithm (conceptual system) that encodes and decodes a bitstring message (conceptual object) are components of a digital conceptual system—namely, a public-key (asymmetric) cryptography system [57].

*Definition 10a: Digitalized Material System (DZMS):*

A *digitalized material system* is an artificial material system that has at least one component that is a digital material object, a digital material system, or another digitalized material system.

For example, a ride-sharing service such as Uber or Lyft is a sociotechnical *digitalized* material system[8] because it is composed of digital material components such as communication and computing devices, user applications, and dispatch and analytics software (all these can be treated as either DMOs or DMSs), along with other artifacts (e.g., conventional cars) and human participants (e.g., drivers, customers, employees). Digitalized systems are not signs per se, as some of their components are not symbols.

*Definition 10b: Digitalized Conceptual System (DZCS):*

A *digitalized conceptual system* is an artificial conceptual system that has at least one component that is a digital conceptual object, a digital conceptual system, or another digitalized conceptual system.

For example, conceptually, to improve the representation and manipulation of data across different components, the designs of computer systems often employ the binary number system in conjunction with the quaternary, octal, and hexadecimal number systems (e.g., [58]). Each of these number systems is a component of an overall conceptual system, which is a numeral position system.

Notably, in much IS discourse when the terms *digital object*, or *digital system*, or *digitalized system* are invoked, it is in the sense of *digital material object, digital material system*, and *digitalized material system* (although these meanings are rarely specified, leading to potential confusion). It is in this sense that concepts such as Enterprise Resource Planning, Generative Artificial Intelligence, social media, database, laptop, hard drive, and others are used—all are material things that have bitstring symbols as components. In contrast, digital conceptual objects, digital conceptual systems, and digitalized conceptual systems are of interest to theoretical computer science and mathematics, where abstract properties of the binary system (including Boolean algebra) are discussed.

[8] Note, not all sociotechnical systems are digitalized material systems. For example, a 19th century mining operation was a nondigitalized sociotechnical system supported by other technologies (e.g., steam engines).

For simplicity, in the remainder of our paper, unless we specify otherwise, we will use the terms digital object, digital system, and digitalized system to mean *things* (material, concrete objects) and not constructs.

### 4.4. Component IV: bitstring symbol mappings and their interpretation

The ways that bitstring symbols map to other objects (constructs or things) and their properties (conceptual properties or real properties) are not straightforward. One reason is that instances of bitstring constructs can be used to represent any object that can be finitely described, computed, or distinguished. A second reason is that the image of the mapping may in turn map to other objects. Most of Bunge [36] is devoted to formalizing the mappings and explaining their sometimes difficult nature. Understanding their nature is important, however, if the nature and functions of bitstring symbols (digital objects and digital systems) are to be understood.

Fig. 3 shows five types of mappings (relations) that exist. Three are direct (solid line), and two are indirect (dotted line). All mappings involve translators. Table 2 shows some examples to illustrate the nature of the mappings.

- *Designation* (Bitstring symbol to Construct): Assigns a meaning to a bitstring symbol via interpretation rules (human or machine).
- *Reference* (Construct to Object): Shows which object (construct or thing) is referenced by the designated construct.
- *Representation* (Construct to Property of Object): Shows which property of a construct or thing is represented by the designated construct.[9]
- *Denotation* (Bitstring symbol to Object): Defined by the relational product of the *designation relation* and the *reference relation* ([36], p. 42).
- *Proxying* (Bitstring symbol to Property of Object): Defined by the relational product of the *designation relation* and the *representation relation* ([36], p. 91).

Often, the relata and relations in Fig. 3 reflect human conventions. In a machine-learning environment, however, Kotlarsky and Oshri [18] point out that the relata and relations may reflect the idiosyncrasies of a particular machine's learning algorithms. The machine may create its own bitstring symbols, constructs, referent objects, represented properties of objects, and mappings. Moreover, the meanings of these machine-created relata and relations may be opaque to humans. The relata and relations may also change frequently as the machine learns.

Note that it is *bitstring symbols* and not *instances of bitstring constructs* that are in the domain of the mappings shown in Fig. 3. In this regard, the *thing* that provides the substrate for the bitstring symbol is important to understanding the meaning of the mapping. For instance, an instance of a bitstring construct that represents the color "red" has a different meaning if it is a property of a sensor (a warning) versus a property of a digitized painting (reflecting the hues of a landscape in a natural setting).

### 4.5. Component V: digital and digitalized systems: hereditary and emergent properties, function, form, and mechanism

As composite things, two types of properties that digital and digitalized systems possess are hereditary properties and emergent properties :[10]

*Definition 11: Digital System, Digitalized System—Hereditary Property:*

A *hereditary property* of a digital or digitalized system is a property in general that is also possessed by at least one of the system's components.

For instance, *processing speed* is a property of a laptop computer as well as some of its components such as its processors and storage devices. The laptop's overall processing speed is some function (perhaps imperfectly known) of the processing speed of its components. Similarly, mean time between failure is a property of a cloud system overall (comprising digital and human components) as well as some of its components (e.g., servers and disks).

*Definition 12: Digital System, Digitalized System—Emergent Property:*

An *emergent property* of a digital or digitalized system is a property in general that is possessed only by the system and not any of its components.

For instance, an emergent property of a social media platform is its potential to impact the mental health of its users [59]. This property is not possessed by any of the platform's components such as the servers used to drive the platform.

Digital objects, digital systems, and digitalized systems perform *functions*. The nature of the concept of "function" is not straightforward. For instance, Mahner and Bunge [60] develop several different concepts of "function" depending on whether the focus is an object's internal processes, external processes (arising from interactions with its environment), both internal and external processes, and adaptive processes. Similarly, after surveying the literature on the concept of "function," Crilly [[61], p. 322] classifies functions as technical, social, and aesthetic.

[9] In the context of Figure 3, we follow Bunge [[38], pp. 69-72] who, for theoretical ease, sometimes treats different values of a property as different properties. Thus, in Figure 3, a construct might *reference* another construct (e.g., a particular construct that is an element of a class of constructs) or a particular thing. The construct also *represents* a particular formal property of the construct *referenced* or a real property of the thing *referenced* where different values of the property would be *represented* (treated) as different properties.

[10] Because we have defined digital objects as atomic bitstring symbols, the classification of properties into hereditary and emergent is not meaningful for them.

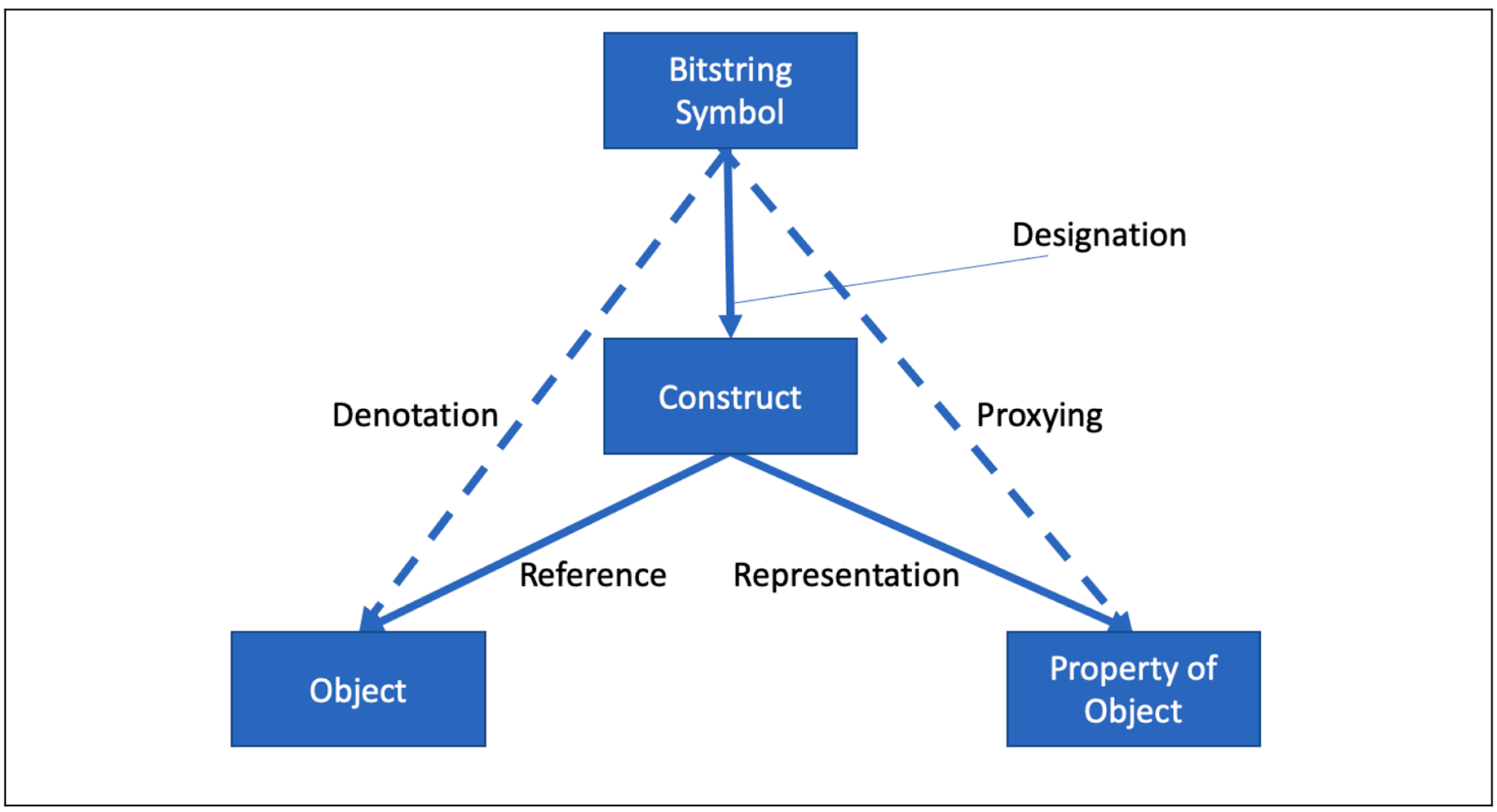

**Fig. 3.** Relationships among bitstring symbols, constructs, and objects (based on [36], p. 43).

**Table 2**
Examples of construct reference mappings and representation mappings.

| Construct | Object(s) *Referenced* | Example Property(ies) of Object *Represented* |
|---|---|---|
| Roman numeral “IV” | Number “4″ | Integer, obeys addition and subtraction rules |
| Sequence of instructions for online sales system | Online sales system | Instructions (transformations) of online sales system |
| “5″ is the cardinality of set *S* | Set *S* | Cardinality of set *S* |
| Birth date field in employee record | Employees | Birth dates of employees |
| Employee with personnel number “342″ | Particular employee | Birth date, salary level, date when employed |
| Cloud storage system icon for display on computer screens | Particular cloud storage system; computer screens | Fees for different storage capacities, encryption methods used |
| Inventory item relation | Inventory items | Attributes of inventory items |

Because our focus is the *intrinsic properties* of digital objects, digital systems, and digitalized systems, we adopt a concept of “function” that is associated only with their *internal* processes:

*Definition 13: Digital Object, Digital System, Digitalized System—Function:*

A *function* of a digital object, digital system, and digitalized system is an outcome (a specific state of the digital object, digital system, and digitalized system) that it is designed to be achieved only via its *internal* processes.[11]

For example, based only on their *internal* processes, a function of a solid-state drive is to store data as bits, a function of a social media platform is to allow people to engage with each other via electronic means, and a function of an ERP system is to represent an enterprise’s states, processes, events, and history.

Functions (in general) are specifically designed *intrinsic* properties of digital objects and *hereditary* properties of digital systems and digitalized systems. Digital systems and digitalized systems have a function, and all of their components have a function. Nonetheless, the *specific function* of a type of digital system or digitalized system will differ from the specific functions of its components. Thus, the *specific functions* of a type of digital system or digitalized system are *emergent properties* of the system.

*Definition 14: Digital Object, Digital System, Digitalized System—Form:*

The *form* of a digital object, digital system, or digitalized system is the set of all the properties that it possesses.

In the extant literature, the concept of “form” is used in different ways. For example, it is sometimes restricted to the component structure of a digital system or digitalized system or the levels of services that they provide. Contrariwise, note that our definition is inclusive. It covers *all* properties, which include the component level structure and relations among properties (e.g., how a function performed by one component depends on a function performed by another component) ([38], pp. 57–109).

[11] Mahner and Bunge [[60], p. 82] would classify our notion of “function” as a teleofunction$_1$, and Crilly [[61], pp. 319, 322-323] would classify it as a physical, technical, proper function.

*Definition 15: Digital Object, Digital System, Digitalized System—Mechanism:*

The *mechanisms* of a digital object, digital system, or digitalized system are the internal processes that occur in them that enable them to enact their function.

For example, a videoconferencing system's mechanisms are the processes that convert pictures and sound to instances of bitstring constructs, transmit them over a communications network, and then translate them back into pictures and sound. These mechanisms allow a videoconferencing system to achieve its function, which is to allow people who are remote from one another to communicate. Similarly, an ERP system's mechanisms are the processes that update instances of bitstring constructs that represent an enterprise's states, processes, and events. The function of an ERP system is to provide an up-to-date representation of an enterprise's past, current, and possible states, processes, events, and history.

Based on our definition of "mechanism," we note the following:

- Mechanisms are processes in *material objects* (things) ([62], p. 21; [60], p. 88). Conceptual systems (constructs) have no mechanism because they have no processes ([62], pp. 24–25).
- The mapping between functions and mechanisms is 1:N ([60], p. 90). For instance, different videoconferencing systems allow their users to communicate with one another, but they may use at least some different mechanisms (e.g., the way they compress instances of bitstring constructs).
- "Every mechanism is a process, but the converse is false" ([62], p. 24). Processes may unfold without apparent organization (e.g., random walk in a stochastic simulation of markets, chaotic spread of spilled ink), whereas mechanisms are processes (often intentionally designed) that enact specific functions of systems.
- Every mechanism enacts change or controls change in a concrete system (thing) ([62], pp. 21, 24).

The functions, form, and mechanisms of a digital object, digital system, or digitalized system are related in some important ways. Specifically, the form of a digital system or digitalized system often follows from the function(s) that they are designed to perform.[12] The form includes the set of properties that a thing must possess if it is to be capable of performing a function.[13] Furthermore, functions only exist (really, not conceptually) when things have mechanisms to enact them.

Also, during the design process, the overall function that is the designed objective for a digital system or digitalized system might be decomposed into smaller functions, which in turn are decomposed further. Because a function is a property of some thing, however, *a decomposition of functions automatically invokes a decomposition of things*—specifically, a *level structure* of things (see Table 1). This resulting level structure shows how the components of the digital system or digitalized system are related. Specifically, each atomic and composite component in the level structure has a function, and it must have a form (its properties, which include a mechanism) that enables it to enact its function. In other words, the overall functions of a digital system or digitalized system cannot be enacted unless *every* function in a decomposition of these functions can be enacted by either a digital object, digital system, or digitalized system. *In short, functions, form, and mechanisms are tied inextricably to their underlying physical substrate.*

## 5. Some implications of our theory

In this section, we show some implications of our theory. While we contend our theory can inform theorizing in many areas relating to digital and digitalized phenomena, we provide three examples in the subsections below to show how it leads to predictions that are contrary to much of the dominant discourse. We also state four broad propositions that our analyses motivate. These in turn can be used as a basis for articulating more specific hypotheses that inform the design and conduct of empirical tests to evaluate the merits of our theory. We provide some examples of hypotheses derived from our propositions and one way that each hypothesis might be tested.

### 5.1. "Emancipation" of form and function from underlying physical substrate

In their analyses of digital systems and digitalized systems, some scholars focus on their function (e.g., the services they provide) and their form (e.g., how they are configured) while they background their underlying material substrate (their level structure of composite things and their components).

For example, Kallinikos [[34], p. 69] argues that the evolution of "technological objects" has led to "a growing emancipation" of their form and function "from the materials in which they are entangled." Similarly, Yoo et al. [[32], p. 726] argue digital innovations are facilitated by two "critical separations": "(1) that between device and service because of reprogrammability and (2) that between network and contents because of the homogenization of data." These "separations" allow digital innovations to take on new functions and forms.

While these arguments are seductive, contrariwise, for three reasons, we argue they are problematic. First, as per our arguments in subSection 4.5 above, function and form are *properties* of things and thus can never be "emancipated" from the *things* they help characterize. They are always real properties of material things.[14] Instead, the so-called "emancipation" that has occurred arises because scientific and technological progress has wrought new things (new forms) or surfaced latent properties of existing things to provide new or improved functionality. This progress manifests as 1:N mappings (with N increasing) between functions and mechanisms ([63], p. 194; [60], p. 90).[15] Similarly, services cannot be "separated" from the devices (systems) that provide them because the services are an *emergent property* of the devices, nor can "contents" or "data" be "separated" from the network (of devices) that possesses the *properties* of being able to store, process, and transmit them.[16]

[12] Sometimes function follows form. For example, when cell phones were first designed and produced with built-in cameras, many users did not know how to fully exploit the camera's functions. Currently, many users are grappling with how to take proper advantage of the AI functionality in their cell phones.

[13] If no such form can be found (e.g., technically impossible or prohibitively expensive), the function cannot be enacted by a thing. Finding an appropriate form is sometimes a major design challenge.

Second, when a digital system's functionality or digitalized system's functionality is compromised, the *mechanisms* that enable the functionality must be investigated. For instance, consider the loss of functionality that occurred in many systems worldwide when the functionality provided by the Crowdstrike system was compromised (e.g., [64]). The loss of functionality quickly became clear to users, many of whom did not even know their systems depended on Crowdstrike. Importantly, the reasons for the loss of functionality could be determined only by individuals who knew the *things* and the *mechanisms* that underpinned the compromised functionality.[17]

Third, a focus on functions and services in digital or digitalized systems is a form of *functionalism* ([60], pp. 84–85).[18] Functionalism is a problematic philosophy because it enables only a superficial understanding of systems ([60], pp. 84–92). A deep understanding can be obtained only by uncovering a system's *mechanisms*. Yet this task can be difficult because mechanisms are often invisible ([62], pp. 27–28; [63], p. 200). Their existence must be hypothesized, and the resulting hypotheses must then be tested. Generating the hypotheses involves careful analysis of the composition of a system, the structure of the system, and the couplings between the system's components and things in the environment of the system ([62], p. 24).

In short, whereas functions denote *what* a thing does, mechanisms *explain* how it does it. When a function fails or is compromised in some way, the underlying mechanisms must be examined, which entails examining the substrate of material components. Accordingly, we propose:

**Proposition 1**. *Designers, managers, users, and researchers who foreground the functionality of digital objects, digital systems, and digitalized systems and background their material substrate and mechanisms will be less able to assess the potential exposures and costs associated with their use.*

The following is an example of a hypothesis that is derived from this proposition:

**Hypothesis 1**. The end users of a hospital health system that relies on the Internet of Things (IoT) will be less able to pinpoint threats to the privacy of patient data when they are presented with a functional architecture of the system versus a component and mechanistic architecture of the system.

The rationale underlying this hypothesis is that an attack on the system can only be done by compromising one of more of its substrate components. While a particular function might be the focus of an attack, the function cannot be compromised except by identifying and undermining the substrate components whose mechanisms enable the function to be performed.

To test this hypothesis, a tabletop exercise[19] with two groups of end users (e.g., doctors, nurses, physiotherapists, and health administrators) of the hospital health system could be employed. One group could be presented with an architecture that shows the functions performed by the system, a description of each function, and how the functions are interconnected. Another group could be presented with an architecture of the system that shows the substrate components of the system, a description of each component's mechanism, and how the components are interconnected. Each group would be asked to identify threats to the privacy of patient data.

Each group's comments would be recorded and later transcribed. Several experts in health system cybersecurity would then read the transcripts, list the threats identified by each group, and rate the materiality of each threat (e.g., see how Jaigirdar et al. [65] used these types of experts). The experts would be blind to which group was the source of a transcript. The transcript for a particular threat would also be randomized by group. Based on Proposition 1, we predict the second group of end users will identify a greater number of threats and a greater number of threats that the expert assessors deem to be more material.

[14] Mahner and Bunge [[60], p. 88] observe that functions "are not substrate-independent. If in doubt, try to build a computer chip with air or iron rather than silicon."

[15] These 1:N mappings underpin equifinality and functional equivalence.

[16] Indeed, the risks of separating data from the material substrate where it is stored, processed, and transmitted become apparent if, as a result, management loses sight of its responsibilities arising from laws about certain types of data: (a) *provenance* (the data is subject to the laws of a jurisdiction regardless of its location; however the location then becomes important to achieving compliance with the laws), (b) *residency* (the data is subject to the laws of the jurisdiction in which it is located), and (c) *localization* (under law, the data *must* be located in a particular jurisdiction) (e.g., [84,85]).

[17] For this reason, Bunge [[63], p. 194] argues that "conflation of 'mechanism' with 'specific function' is not advisable when one and the same task can be performed by different mechanisms—the cases of functional equivalence."

[18] "Functionalism is usually understood as the (ontological) thesis that function is all-important and stuff (or composition) nothing; more carefully stated, stuff is relevant at most as the material carrier of functions, but, inasmuch as two things perform the same function$_{1,2}$, their material or compositional differences do not matter" ([60], p. 84).

[19] A tabletop exercise is an empirical research method now often used in cybersecurity research (e.g., [86,65,87]). The participants play a role, and they respond to various scenarios presented to them by a facilitator. The facilitator also keeps the participants on task. Tabletop exercises are akin to focus group exercises (e.g., [88]).

### 5.2. Problematic quasi-reification of bitstring constructs

In our theory, recall that we distinguish between *bitstring constructs* and *bitstring symbols* (Definitions 5 and 7 above). The former are *nonmaterial* objects and possess *formal* properties. The latter are *material* objects and possess *real* (material) properties. Contrary to our theory, however, some scholars quasi-reify bitstring constructs by assigning *real* properties to *nonmaterial objects*.[20]

For instance, Faulkner and Runde [[4], p. 1285] deem bitstrings to be syntactic objects, which are *nonmaterial* objects in their theory. However, in relation to the properties of bitstrings, they say: "[o]ne such property is its nonmaterial mode of being, from which flow others such as its not degrading with repeated use and what economists call the property of non-rivalry that its use by one person in no way impinges on its simultaneous use by any number of others" ([4], p. 1286).

If Faulkner and Runde's focus is *bitstring constructs*, we agree that bitstring constructs possess the *formal* properties of "non-degrading" and "non-rivalry." These are properties of many, perhaps all, *constructs*. For instance, Boolean algebras do not degrade with repeated use, nor is their use by one mathematician depriving another mathematician from using them.

However, Faulkner and Runde's narrative reveals their focus is a mixture of bitstring constructs and bitstring symbols. For instance, to operate on nonmaterial objects, they say: "[a] basic feature of nonmaterial objects is that in order to be accessed—to be used, stored, passed to others, etc.—they must in some way be inscribed onto, contained within, or borne by a material object of some kind" ([4], p. 1285). We contend that their concepts of "use," "being stored," and "being passed onto others" are *real* properties, which they have assigned to a nonmaterial object (they make little sense as *formal* properties).[21] This apparent anomaly is perhaps the reason why Faulkner and Runde [[4], p. 1286] say that "a material bearer is necessarily a hybrid object," which are "objects that comprise both material and nonmaterial objects as component parts" ([4], p. 1284).

Moreover, whereas bitstring *constructs* might not degrade and permit simultaneous use, bitstring *symbols* indeed degrade with repeated use (e.g., a disk fails and loses its stored bitstrings), and use by one object precludes its simultaneous use by another object (e.g., two different programs cannot process their bitstrings using the same core in a chip at exactly the same time). Moreover, it is *bitstring symbols*, and not *instances of a bitstring construct*, that are used by digital objects, digital systems, and digitalized systems to enact their states, events, and processes.

Fig. 4 below shows our alternative representation of Faulkner and Runde's [[4], p. 1289] Fig. 3, where seemingly real properties characterize nonmaterial objects that in turn are inscribed onto material objects. They use their diagram to show that nonmaterial "bearers" can be "bearers" of other nonmaterial objects. For example, a Zip file is a nonmaterial bearer of a nonmaterial DOCX file. At the same time, a DOCX file can be translated into a Zip file (and vice versa). Nonetheless, ultimately, each nonmaterial bearer must be inscribed on a material substrate.

Contrariwise, our theory eschews convoluting nonmaterial objects with material objects and thus it denies the existence of hybrid objects. Fig. 5 below shows how we would represent Faulkner and Runde's Fig. 3 "layers" of bearers according to our realist theory of digital objects.

By convoluting bitstring constructs and bitstring symbols, we argue two pitfalls arise:

*1. Exposures associated with the materiality of bitstring symbols are overlooked*

When bitstrings are quasi-reified, we predict that designers, managers, users, and researchers who engage with digital objects, digital systems, and digitalized systems may systematically ignore or forget about the materiality of bitstring symbols. For example, if the thing (bearer) is at some unknown location in the cloud, potentially it can be compromised by a hostile party, act unreliably, and so on. Absent an untoward experience with a bearer (as per the Crowdstrike failure discussed above), many important questions about bitstring symbols may not be asked —for instance, the extent to which they are reliable, protected, durable, environmentally friendly, and accessible (see, e.g., [66], pp. 220–226). Accordingly, we predict:

**Proposition 2**. *Designers, managers, users, and researchers who engage with digital objects, digital systems, and digitalized systems and who focus primarily on bitstring constructs rather than bitstring symbols will be less attuned to the potential exposures and costs associated with their use*.

The following is an example of a hypothesis that is derived from this proposition:

**Hypothesis 2**. The end users of a large language model (LLM) that has been developed to support design work in creative industries will be less able to identify threats arising from intellectual property infringements[22] when they are presented with a description of the knowledge bases on which the LLM has been trained versus the material substrates that map to the knowledge bases on which the LLM has been trained.

The rationale underlying this hypothesis is that end users must be able to evaluate the *provenance* of the material used to train the LLM if they are to be better able to evaluate whether the output is likely to contain intellectual property violations. Essentially, a named source knowledge base is a bitstring construct (a sequence of bits); however, it might have been assembled from multiple material substrate components. Each material substrate component (e.g., a physical database and the computer system that maintains it) is a bitstring symbol. It must be evaluated to determine the risk of it containing intellectual property violations.

[20] Hence, our use of the term "quasi-reified" because the constructs are deemed nonmaterial but their properties are treated (sometimes implicitly) as real.

[21] At first glance, constructs can also be "used" and "passed onto others." But of course, these actions can only be done by a brain. So, ultimately, they are real properties of a brain.

[22] For instance, infringements occur because a particular knowledge base contains material that violates copyright, trademark, patent, and registered design laws.

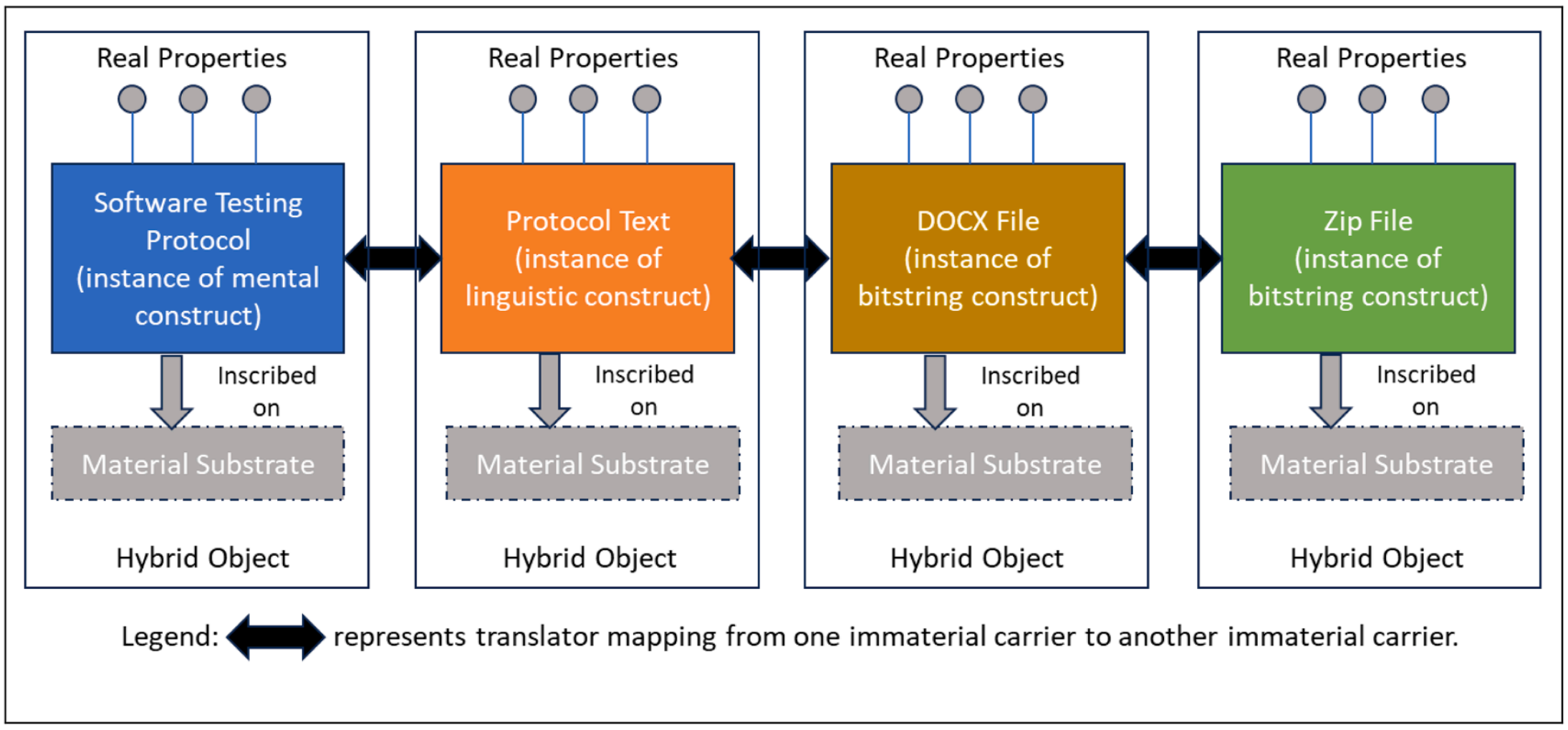

**Fig. 4.** Alternative representation of Faulkner and Runde's [[4], p. 1289] Fig. 3, which is an example of layers of nonmaterial and material bearers.

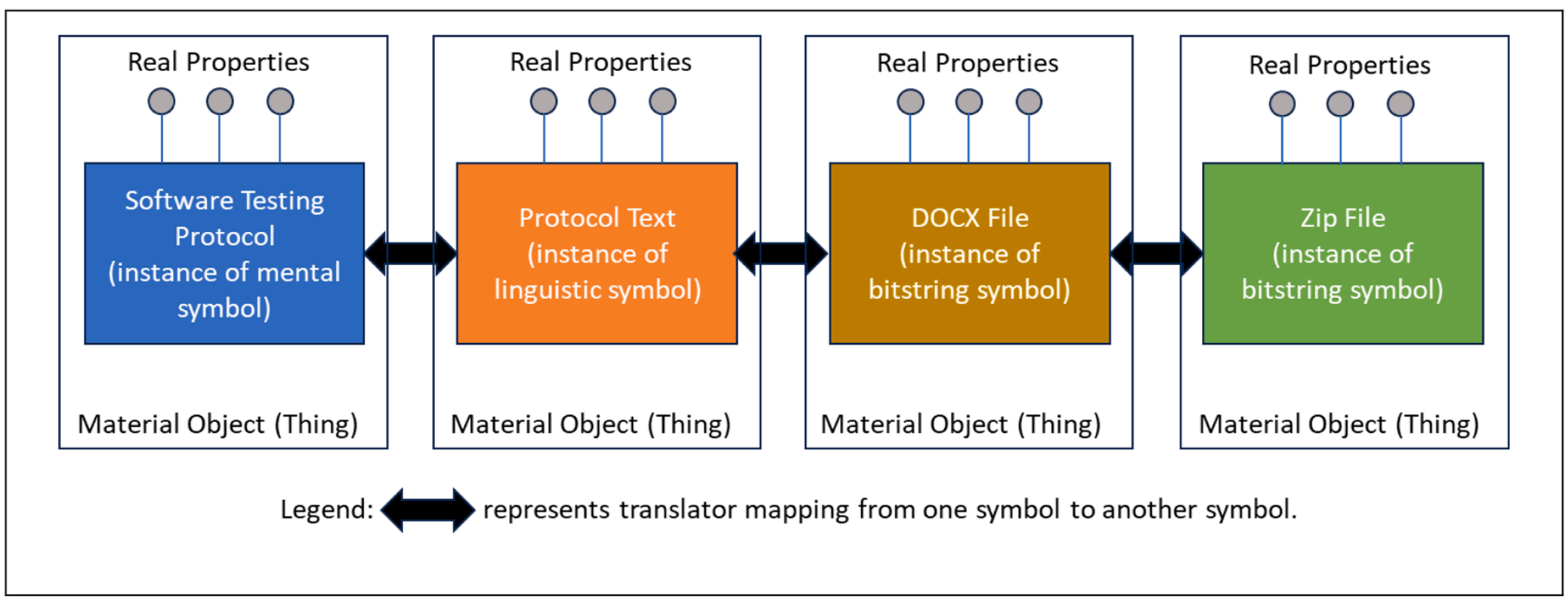

**Fig. 5.** Representation of Faulkner and Runde's [[4], p. 1289] layers of bearers using our realist theory of digital objects.

To test this hypothesis, two groups of end users of the LLM could be employed in a tabletop exercise (see Hypothesis 1 above). One group could be presented with a description of each knowledge base used to train the LLM. Another group could be presented with a description of each knowledge base and the material substrate components from which each knowledge base has been assembled.[23] Each group would be asked to identify possible threats arising from the output of the LLM containing intellectual property infringements.

Several experts in cybersecurity, provenance associated with the domain in which the LLM operates, and intellectual property law could be used to read transcripts of each group's comments. The experts would be blind to which group was the source of a transcript. The transcript for a particular threat would also be randomized by group. The experts would be asked to list the threats identified by each group and to rate the materiality of each threat. Based on Proposition 2, we predict the second group of end users will identify a greater number of threats and a greater number of threats that the expert group deems more material.

*2. Failing to recognize the occurrence of material substrate changes and their implications*

Unbeknown to many stakeholders, a bearer of bitstrings might be changed, perhaps to a thing that is cheaper, more unreliable, and

[23] So the two groups receive roughly the same amount of materials, the first group could also be given some "distractor" materials, such as a description of the ways LLMs are being used in creative industries. The downside of using distractor materials, however, is that they might undermine the first group's ability to perform the experimental task.

slower, or a thing that is more expensive, more reliable, and faster but a higher polluter. When properties are "realized" in different things, ripple effects can occur. For instance, the *emergent properties* of a digitalized system that has a change in a component may alter unexpectedly (e.g., its overall robustness). As a result, website users might find they can no longer perform certain functions or a trapdoor suddenly appears that exposes the system to actions by malicious parties. Accordingly, we predict:

**Proposition 3**. *Designers, managers, users, and researchers who engage with digitalized systems and who focus on bitstring constructs and not bitstring symbols will be less able to explain how changes to a digitalized system's component things affect how it operates*.

The following is an example of a hypothesis that is derived from this proposition:

**Hypothesis 3**. The end users of a large language model (LLM) that has been developed to support how a political party gauges public opinion on different matters will be less able to identify where cyberattacks that introduce biases into data sets used to fine-tune the LLM are occurring when they are presented with a description of these data sets versus the material substrates that map to the data sets.[24]

The rationale underlying this hypothesis is that end users must be able to identify the possible *physical locations* (bitstring symbols) where a cyberattack that introduced a specific bias in the LLM's output is initiated. Pinpointing a training set as a possible source of bias is insufficient as a means of identifying the location of a possible cyberattack because a training data set may have been assembled from multiple material substrate sources.

Using a tabletop exercise, two groups of individuals who have experience with LLMs and are active politically could be presented with a description of a list of biases that have been identified in an LLM's output. One group would also be presented with a description of the data sets used to fine-tune the LLM. Another group would be presented with a description of the fine-tuning data sets and the material substrate sources from which each fine-tuning data set has been assembled. For each bias, each group would be asked to indicate the location of and how they believe one or more training sets has been compromised by a cyberattack. Each group's comments would be recorded and transcribed.

Several experts in cybersecurity and LLMs would be used to read transcripts of each group's comments. The experts would be blind to which group was the source of a transcript. For each bias, the order in which a transcript is presented to the experts would also be randomized. The experts would be asked to evaluate the plausibility and location of possible cyberattacks given by the group for a specific bias. Based on Proposition 3, we predict the experts will deem the second group better able to identify plausible locations and the nature of possible cyberattacks that have led to a particular bias in the LLM's output.

### *5.3. Layered modular architecture*

Yoo et al. [[32], p. 724] argue that "pervasive digitization gives birth to a new type of product architecture: *the layered modular architecture*." They contend (p. 724) that a "layered modular architecture *extends the modular architecture of physical products* by incorporating four loosely coupled layers of devices, networks, services, and contents created by digital technology" (our emphasis). More recently, the layered modular architecture has been extended to cover "data" ([1], p. 1511), if not in the form of a separate layer then in the form of the "homogenization" of data as part of the "contents layer."

Based on our theory, we do not see that "pervasive digitization" has given birth to a new type of product architecture. Instead, we see the evolution of digital objects and their widespread diffusion and infusion into various other devices has led to new *systems* that possess emergent properties not possessed by prior systems. We simply have a level structure of components (things), which is a property of *all* systems (Table 1). Moreover, as per our arguments above, "services" and "contents" are *not* things. Instead, they are *properties* of things—namely, properties of digital objects, digital systems, and digitalized systems (they cannot exist independently of these things).[25] Indeed, services and contents have *always* been properties of digital objects, digital systems, and digitalized systems.[26]

By conceiving of services and contents (properties) in the same way as devices and networks (things), we contend several problems arise. First, the underlying architecture of the system is masked. Where services and contents are *emergent* properties, the ways in which the component digital objects, digital systems, and digitalized systems are coupled need to be understood if the architecture that underpins the delivery of the services and contents is to be understood.[27] Moreover, we predict that clearly understanding the ways in which the component digital objects, digital subsystems, and digitalized subsystems of a digitalized system are coupled is fundamental to being able to articulate the *mechanisms* that underlie a digitalized system.

Second, we disagree that a layered architecture manifests "…two critical separations: (1) that between device and service because of reprogrammability and (2) that between network and contents because of the homogenization of data" ([32], p. 726).[28] Properties of things cannot be separated from the things they characterize—thingless properties do not exist ([38], p. 64). A digital object's or digital system's or digitalized system's property of "reprogrammability" simply means it can take on new properties. Similarly, a network's property of "homogenized data" simply means some components of the network have the property that they can represent phenomena digitally.[29]

[24] See Ferrara [89] for a definition of and the types of "bias" in large LLMs and a discussion of the ways in which the choices of training data and use-cases can effect biases in LLMs.

[25] In the context of our theory, in their Figure 1, "The Layered Architecture of Digital Technology," Yoo et al. [[32], p. 727] convolute functionalities (contents and services), which are properties of things, with the things that provide the functionalities (devices and networks).

[26] Yoo et al. [32]) rightly point out, however, that the services and contents they now provide continue to change rapidly and have increasing utility (in other words, new emergent properties manifest in these new systems).

[27] Indeed, our view is consistent with Yoo et al.'s [32] definition of "digital innovation"—namely, " the carrying out of new combinations of digital and physical components to produce novel products."

[28] We argue these properties of digital objects and digital systems are not new—they have always existed.

Third, because "reprogrammability" and "homogenized data" are properties of things, they cannot beget properties ([38], pp. 98–99). By reifying these two properties, we hypothesize that pinpointing the full set of emergent properties that have arisen through couplings among digital objects, digital systems, and digitalized systems will be inhibited. The focus will be the *known set* of services and contents and not services and contents that initially are invisible or can be created through novel couplings among digital objects, digital systems, and digitalized systems. Moreover, predicting new emergent properties that might arise through new forms of coupling among digital objects, digital systems, and digitalized systems will be undermined because possible interactions among them will be opaque or only translucent.

Fourth, a digital innovation strategy ultimately depends on whether digital objects, digital systems, and digitalized systems can be coupled to form systems that possess new properties—specifically, properties that provide new functionalities the marketplace will demand. If digital objects, digital systems, and digitalized systems cannot be coupled in the first place, emergent properties cannot be created deliberately as part of a firm's strategy or formed synergistically through the actions of others ([32], pp. 729–730).

In light of our arguments above, we predict:

**Proposition 4.** *By focusing on the level structure of components (things) in a digitalized system rather than a "layered modular architecture" of devices, networks, services, and contents, designers, managers, users, and researchers will be better able to identify and exploit the emergent properties (functions, services) that arise in each component of the level structure.*

The following is an example of a hypothesis that is derived from this proposition:

**Hypothesis 4.** Compared to junior engineers, senior software engineers will be better able to pinpoint where functions that are emergent properties in a digital system are located and describe the system in terms of a level structure of components and rather than a layered modular structure of devices, networks, services, and contents.

The rationale underlying this hypothesis is that system designers must be able to discern how the possible emergent properties of a composite (e.g., a function performed by the composite) could arise (or be designed) from the properties of its components (and thus is feasible for the composite to perform). In the absence of a level structure of composites and components, the relationships among emergent properties and hereditary properties are opaque.

This hypothesis could be evaluated using qualitative interviews with software engineers who are tasked with maintaining a digital system (e.g., debugging the code, making periodic code updates, managing cybersecurity of the system). The interviews will be conducted with two groups of software engineers: (a) junior engineers (e.g., one year or less on the job), and (b) senior engineers (over 10 years of relevant experience, reflecting a common threshold of ten years to become a domain expert).[30]

Each interview will focus on the specific digital system the engineer is tasked to maintain. Engineers will be first asked to describe what training material was presented to them about the system, followed by open-ended questions about the system itself (e.g., how the system is structured, where specific system functions are located, and how new functions or capabilities might be introduced). Interview transcripts will be analysed to assess the vocabulary employed (e.g., independent and layered versus level-based component interactions).

Consistent with Proposition 4, we expect systematic differences will be evident across experience levels. Specifically, junior engineers are expected to rely more heavily on layered modular descriptions, referring to devices, data, or network layers as independent entities. Moreover, any explanations of emergent functionalities they provide are expected to be vague and disconnected from the underlying components and component interactions. By contrast, senior engineers are more likely to locate functions as emergent properties of specific composites and to explain how changes in component interactions could give rise to new or previously unrealized functionalities.

In short, based on Proposition 4, we expect the interviews to reveal distinct patterns of reasoning among senior engineers because they focus more on level-structured explanations and exhibit greater consideration of functional emergence due to component interactions. As a. result, they have greater confidence in their ability to understand and manage the functions of the target systems.

## 6. Conclusions

In this paper, we have proposed a realist theory of digital objects, digital systems, and digitalized systems to clarify their nature and provide a basis for future theorizing and empirical work. Our theory is summarized in Appendix B, which contains the definitions we have articulated above of the constructs that provide the scaffold for our theory.

[29] Yoo et al. [[32], p. 726] also argue that a third "key characteristic" of digital innovation is its "self-referential nature," which refers to the need for one form of digital technology needing to reference another form of digital technology," thereby leading to network externalities. In the context of our theory, these network externalities are an emergent property of the underlying system, which is comprised of digital objects and digital systems.

[30] We recognize the problematic nature of this threshold, but nonetheless it is a useful, initial rule-of-thumb (e.g., https://www.verywellmind.com/expertise-how-hard-is-it-to-become-an-expert-at-something-4173614).

Our theory is at odds with much of the prevailing discourse that characterizes digital phenomena in terms of combinations of material objects, nonmaterial objects, and hybrid objects. Instead, our theory is based on a rigorous philosophy of scientific realism developed by Bunge (e.g., [47,67]) that countenances only material objects in the world.

We believe that debates about the merits of different theories tend to be counterproductive, however. For this reason, we have articulated four propositions that could be used to formulate hypotheses and to design and conduct empirical tests of our theory. Our theory can also be used to analyze other narratives about digital objects, digital systems, and digitalized systems, especially those founded on an ontology that incorporates nonmaterial and hybrid objects or separates the properties of digital objects, digital systems, and digitalized systems from their material substrate. These analyses should enable additional propositions to be developed as a basis for testing the merits of our theory. If our theory is to gain acceptance by other scholars, we acknowledge that (a) these propositions must be *distinctive* (relative to those generated on the basis of current narratives about digital phenomena), and (b) they must withstand "demanding empirical corroboration" ([67], p. 97). Importantly, empirical tests will surface the relative power of two competing theories, which at first glance might not be apparent.

Finally, contrary to arguments made by some scholars, we contend we do *not* need a fundamentally new theory (e.g., [3]) or an "ontological reversal" (e.g., [2]) to understand the nature of digital objects and digital systems because they are a type of sign—specifically, a non-iconic artificial sign (symbol). Thus, they are part of a semiotic system, and semiotic systems have been studied extensively [18,19,40,54,68–72]. Nonetheless, our theory shows that understanding why and how digital objects, digital systems, and digital components of digitalized systems are particular types of symbols (bitstring symbols) is not straightforward.

## Funding

This research received no specific grant from public, commercial, or not-for-profit funding agencies.

## CRediT authorship contribution statement

**Roman Lukyanenko:** Writing – review & editing, Writing – original draft, Formal analysis, Conceptualization. **Ron Weber:** Writing – review & editing, Writing – original draft, Conceptualization.

## Declaration of competing interest

The authors declare that they have no known competing financial interests or personal relationships that could have appeared to influence the work reported in this paper.

## APPENDIX A

*Summary of Key Prior Research that Informs Narratives on the Nature of Digital Objects, Digital Systems, and Digitalized Systems*[1]

| Authors | Research Focus | Definition of Digital Object, Digital System, and Digitalized System | Attributes/Properties of Digital Object, Digital System, and Digitalized System | Some Implications |
|---|---|---|---|---|
| Kallinikos [73] | Implications of using digital technology to create representations of contexts increasingly separated by time and space. | Not defined. Focus is digital technologies. | Not identified. Focus is on the nature of representations enabled by digital technologies. | Digital technologies allow actions to be coordinated across contexts separated in space and time through their ability to represent different contexts. |
| Buckland [74] | Nature of documents, especially digital documents. | Documents, including digital documents, are "signifying objects." | Properties of documents, including digital documents:<br>• materiality<br>• intentionality (to be used as evidence)<br>• outcome of a process<br>• perceived as a document | In a digital environment, the distinctiveness of documents as a physical form is undermined. |
| Manovich [26] | The nature ("language") of new media (computer-based), especially in contrast to cinema and photography. | The term "new media object" (e.g., digital image, virtual 3D environment, web page) is preferred over digital object as previous media types (e.g., cinema) were also partially digital. All new media objects "are composed of digital | Principles of new media are:<br>• numeric representation<br>• modularity<br>• automation<br>• variability<br>• transcoding | Broad cultural implications are discussed, with the suggestion of the emergence of new "information culture." |

(*continued on next page*)

(*continued*)

| Authors | Research Focus | Definition of Digital Object, Digital System, and Digitalized System | Attributes/Properties of Digital Object, Digital System, and Digitalized System | Some Implications |
|---|---|---|---|---|
| | | code; they are numerical representations" of physical objects. | | |
| Thibodeau [27] | Preservation of digital objects. | "A digital object is an information object, of any type of information or any format, that is expressed in digital form." p. 6. | Digital objects have three types of properties:<br>• physical properties<br>• logical properties<br>• conceptual properties | Methods for preserving digital objects must be feasible, sustainable, practicable, and appropriate. Digital preservation involves a paradox—deliver past digital objects authentically but use the latest technology. |
| Allison et al. [15] | Nature of "identity" with digital objects. | A digital object is a bitstring encoded in a particular format that communicates information. | • Identity of digital objects.<br>• Equivalence of digital objects.<br>• Transitivity of digital objects. | The same bitstring may be stored in different formats and use different hardware and software. Thus, its identity is problematic. Also, the experience of digital objects (bitstrings) for humans is not fixed. It is mediated by the hardware and software that are used. The same bitstring may produce different experiences. |
| Benkler [24] | Based on the proliferation of "connected personal computers," characterize a new type of economy— "networked information economy" | Digital object is conceptualized as a connected personal computer—a device for processing, storing, and communicating information. | Connected personal computer has two key capacities: "to make meaning—to encode and decode humanly meaningful statements," and "to communicate one's meaning around the world." | Enhanced autonomy of individuals; emergence of new public sphere and decentralized discourse; positive impact on justice and human development; emergence of more critical and self-reflective culture. |
| Lessig [25] | Understanding the nature of code and cyberspace, how cyberspace presently evolves and how it could evolve. | Digital objects are not defined; rather, the focus is cyberspace and code. Code is "the instructions embedded in the software and hardware that makes cyberspace what it is." Cyberspace is the product of code. | Cyberspace is created and governed by code, whereas "real space" is governed by laws of humans and nature. Code is created by humans based on changing priorities and values. | Legal, political, and commercial implications of different cyberspace architectures for regulation, jurisdiction, and sovereignty as well as intellectual property, privacy, and free speech. |
| Zittrain [75] | Understanding the nature, benefits, and limitations of generative information technologies versus "information appliances." | Digital objects fall into two principal types: *generative* (flexible and extensible by its users, e.g., PCs, Internet, Wikipedia); and *information appliances* (those tailored to a specific task or context, e.g., iPhone, Xbox gaming console). | Generative information technologies have the properties of leverage, adaptability, ease of mastery, accessibility, and transferability. Information appliances are functionally rigid but more predictable and reliable. | Whether digital objects are primarily generative or appliances impacts technological and business innovation, cybersecurity, privacy, software development (e.g., use and impact of open-source software) and level of involvement of people with digitalized systems. |
| Ekbia [3] | Develop a theory of digital artifacts to understand their nature. | A digital object is a quasi-object—a mediator that evolves through the interactions of stakeholders. | • Objects are mediators and not intermediaries because they shape meaning.<br>• Poor "fixity" of digital objects.<br>• Immanent, unstable, hybrid, material<br>• Lacking clear identity<br>• Performative rather than representing something | The nature of a digital artifact emerges through interactions. Different "observers" of a digital artifact may ascribe different meaning to it. |
| Faulkner and Runde [29] | Identity of a technological object. | Not defined. Focus is technological objects. | • Identity<br>• Form | • Nature of technological objects is constituted by |

(*continued on next page*)

(*continued*)

| Authors | Research Focus | Definition of Digital Object, Digital System, and Digitalized System | Attributes/Properties of Digital Object, Digital System, and Digitalized System | Some Implications |
|---|---|---|---|---|
| | | | • Functions | their physical form and social function.<br>• Physical form of technological objects underdetermines their functions.<br>• Functions are assigned to technological objects by social groups.<br>• Identities of technological objects a real feature of the social world. |
| Kallinikos [76] | How technology of computation shapes and reshapes the ways humans perceive the world. | Technological and computational objects are not defined but appear to be similar to, if not the same as, digital objects. | • Computational objects are vertically stratified.<br>• Computation is performative. | Binary nature of computation enables exchanges to occur across qualitatively different domains of “natural, social, and technical reality.” |
| Kallinikos, Aaltonen, and Marton [77] | “Provide the conceptual scaffold on which to develop a theory of digital objects.” | • Digital objects include all digital technologies, digital devices, and digital cultural artifacts (e.g., music).<br>• “Digital objects are objects only in an elusive and perhaps euphemistic way.”<br>• Digital objects are “marked” by a limited number of attributes. | • Editability.<br>• Interactivity<br>• Openness<br>• Distributedness<br>• Modular composition<br>• Granularity | Digital objects are constantly changing as a result of the ways social practices engage with their attributes. For instance, in a digital environment, what constitutes a document to be archived? |
| Leonardi [31] | The nature and different senses of digital materiality in social studies. Three senses of material: touchable—tangible “stuff”; material as performative—providing people with capabilities to accomplish goals; material as significant—important consequences. | Data and electricity are not objects—they are “stuff” without tangible character. You can’t touch data. If material is defined as having physical matter, software is not material. Software exploits some type of social practice that compels people to follow the abstract plan. | Material properties offer certain opportunities and constraints that simply cannot be overcome. Conceptual properties offer a particular freedom of improvisation and re-creation that the physical does not. | Social science research has a unique opportunity to focus on material as practical instantiations and importance. Instead of insisting that it is matter that matters, what may matter most about “materiality” is that artifacts and their consequences are created and shaped through interaction. |
| Yoo [22] | Experiential computing; computing in everyday life experiences. Focus is digital artifacts. | No definition given of digitalized artifacts. However, digital materiality is said to be different from physical materiality. | Properties of digital artifacts and infrastructure:<br>• Programmability<br>• Addressability<br>• Sensibility<br>• Communicability<br>• Memorizability<br>• Traceability<br>• Associability<br>Digital materiality is:<br>• fungible<br>• ephemeral<br>• indeterministic<br>Physical materiality is:<br>• rigid<br>• stable<br>• tangible | Digital artifacts increasingly mediate human experience of the world at the individual, group, organizational, and community levels. |
| Yoo et al. [32] | New product architectures arising as a result of digitization. | No definition given of digital objects. “Digitization” is defined as “encoding of analog information into digital format.” “Digital innovation” is defined as using new combinations of digital and physical components to produce novel products.” | “Key characteristics” of digital innovation:<br>• Reprogrammability<br>• Homogenization of data<br>• Self-referencing | Digital artifacts are embedded into layered, modular architectures that help separate content from devices and information infrastructures. The outcome is “profound changes in a firm’s |

(*continued on next page*)

(*continued*)

| Authors | Research Focus | Definition of Digital Object, Digital System, and Digitalized System | Attributes/Properties of Digital Object, Digital System, and Digitalized System | Some Implications |
|---|---|---|---|---|
| | | | | organizing logic and innovation." |
| Kallinikos and Mariátegui [78] | Understanding the processes by which video production is carried out as digital technologies evolve. | Digital objects are an "ensemble of operations temporarily stabilized." | Digital objects are:<br>• Open<br>• Editable<br>• Expandable<br>• Composed of fine-grained operations<br>• Networked | Video metadata is increasingly important as a means of coordinating activities across multiple people, locations, and platforms as digitization allows video production to be unbundled and broken up into smaller components and processes. |
| Hui [79] | To contrast digital objects with past conceptualizations of natural and technical objects. | Digital objects are objects on the World Wide Web that are composed of data that is formalized via schemas or ontologies that generalize as metadata. | • Programmable<br>• Relational<br>• Changeable<br>Digital objects appear in three phases:<br>• objects<br>• data<br>• networks | • "Bits are the atomic state of representation."<br>• "Temporal state of information is a digital information process."<br>• "Investigation of digital objects to find a new relation between object and mind." |
| Kallinikos [34] | To show that "technological evolution has loosened the bonds between functions, form, and matter" and that function has become progressively more important. | Not defined. Reference made to "digital world, picture, images, ecosystem" etc. Nature of digital object assumed to be known, | Not defined. | Technological advances have resulted in the "emancipation" of form and function "from the materials in which they are entangled." |
| Yoo [21] | To explore how the materiality of digital artifacts enables their generative evolution. | Digital objects and digital systems not defined. However, "digitalization" (a process) is defined as (a) encoding analog information in a digital format, and (b) possible subsequent reconfigurations of the socio-technical context in which digital products are consumed and used. | Four attributes of digital technology:<br>• all information can be structured as binary digits<br>• digitalization embedded into non-digital tools<br>• has immaterial attributes<br>• an enabling technology | • Evolution of digital artifacts can be studied using a genetics-based approach.<br>• Digital artifacts can be characterized as a sequence of a set of fixed base elements or "mutations" to these fixed base elements. |
| Faulkner and Runde [80] | To propose an ontological foundation and a general theory of technological objects within critical realism's Transformational Model of Social Reality. | • "The defining characteristic of objects is that are structured continuants"—composed of parts that are always present throughout the object's existence.<br>• A technological object is one "that has one or more uses assigned to it by the members of some human community."<br>• Two classes of technological objects: material and nonmaterial. | Properties of technological objects:<br>• Identity (based on function and form)<br>• Social position<br>Properties of nonmaterial technological objects:<br>• Non-rivalry in use<br>• Non-excludability | • Additional material bearers of nonmaterial technological objects can be created almost instantaneously at almost zero cost.<br>• Non-excludability property of nonmaterial technological objects creates problems for protection of intellectual property rights. |
| Kallinikos et al. [5] | "To shed light of the conditions in which digital artifacts are currently embedded." | Digital artifacts are objects "that lack the plenitude and stability offered by traditional objects and devices." | Digital artifacts are:<br>• Intentionally incomplete<br>• Perpetually in the making<br>• Modular<br>• Granular in constitution<br>• Editable<br>• Interactive<br>• Reprogrammable<br>• Distributable<br>• Functionally agnostic | • Digital artifacts challenge two fundamental principles of archival practice: provenance and authenticity.<br>• Digital objects introduce a double instability into information search and retrieval as both the target and displayed content from search engines constantly adapt to each other.<br>• Digital objects are embedded in larger |

(*continued on next page*)

(*continued*)

| Authors | Research Focus | Definition of Digital Object, Digital System, and Digitalized System | Attributes/Properties of Digital Object, Digital System, and Digitalized System | Some Implications |
|---|---|---|---|---|
| | | | | digital ecosystems where relationships among objects in these ecosystems constantly change. |
| Brynjolfsson and McAfee [7] | Impact of rapidly increasing amounts of digital information. | Digitization is turning all kinds of information and media into bits. | Economic properties of digital information:<br>• use is non-rival<br>• low cost to reproduce | Increased amounts of digitalized information facilitate better understanding and prediction of how the world works. |
| Hui [81] | To investigate the nature of and existence of digital objects as mediated through their production, implementation, and use. | Focus is limited to representation and categorization of *data* digital objects, which are objects “that take shape on a screen or hide in the back end of a computer program, composed of data and metadata regulated by structures or schemas.” | • Concrete<br>• Universal<br>• Interoperable<br>• Extensible<br>• Recursive<br>• Discursive relational<br>• Programmable | • Digital objects form a “digital milieu” through their relations with other objects.<br>• What happens when humans and machines become reticulated through digital social networks? |
| Baskerville et al. [2] | Ways in which digital representations increasingly are created first and then physical representations. | Use Hui’s definition (see row above). | Two essential properties of digital objects:<br>• nonmateriality<br>• computed nature | An “ontological reversal” has occurred in the sense that information now not only reflects reality but creates and shapes reality. |
| Bogea Gomes et al. [16] | What is digital technology? | Digital technology is a “type of technology involving one or more digital objects whose components include one or more bitstrings.” | As per Faulkner and Runde [4]. | Using the Unified Foundational Ontology [43], builds a conceptual model of Faulkner and Runde’s [4] theory “to provide conceptual clarification and unambiguous communication.” |
| Alaimo and Kallinikos [23] | Theorize the relationship between data, knowledge and organizations | Data is defined as semiotic artifact, instrument of knowing used to capture, represent, know, and act upon the world. *Data object* is a cluster of individual data items. Data objects furnish a layer of meaningful entities upon which further and often elaborate operations are built | Digital data objects are:<br>• content-agnostic<br>• nonneutral<br>• homogenizing<br>Dependency and entanglement with the materiality of the language and machines used to record digital objects. | The entanglement of digital technologies with data carries profound implications for organizations, including unbundling of data with knowledge objects and domain knowledge, resulting in decentering of organizations. |
| Grover and Lyytinen [17] | The information systems field needs new forms of indigenous theorizing if it is to deal adequately with digital phenomena. | Not defined. However, all digitalization operates with bitstrings. | • Bitstrings embody both material and non-material objects.<br>• Digital technologies are combinatorial, relational, and constantly evolving and converging. | Digital technologies have:<br>• produced unprecedented changes in scale and scope of operations;<br>• opened unprecedented opportunities in design;<br>• fostered unprecedented connections across the material and social world;<br>• created multi-level, complex, dynamic socio-technical change;<br>• offered infinitesimal marginal costs with nearly unlimited scalability. |
| Kotlarsky and Oshri [18] | Examines the meaning of digital objects using a semiotics lens and the impact of artificial intelligence on interpreting | Use Faulkner and Runde’s [4] definitions. | With developments in artificial intelligence, the attributes of digital objects are frequently changing, often in | • Digital objects are denoted by signs, but these signs are interpreted to ascribe |

(*continued on next page*)

(*continued*)

| Authors | Research Focus | Definition of Digital Object, Digital System, and Digitalized System | Attributes/Properties of Digital Object, Digital System, and Digitalized System | Some Implications |
|---|---|---|---|---|
| | Faulkner and Runde's [4] theory of digital objects through a semiotics lens. | | ways not understood by humans. | meaning to a digital object.<br>• Developments in artificial intelligence highlight that the signs that denote digital objects, the ways these signs are interpreted, and the digital objects themselves are constantly changing. |
| Winkelnkemper and Schulte [28] | Create an ontology of digital artefacts that show their potential uses. | Not defined. | A large number of properties that fundamentally support:<br>• interactive processing<br>• virtuality<br>• transparent networking | Can be used to describe the digital world without relying on the substantive knowledge of computer scientists and engineers. |
| Yoo et al. [1] | From multiple perspectives, examines the current status of digital innovations and the reasons for the positive and negative consequences that have occurred. | Not defined. References definitions given in earlier work by, for example, Faulkner and Runde [4] and Yoo et al. [32]. However, the term "digital" is now used to "signal the emergence of new technology in organizations." | References types of attributes identified in earlier work by, for example, Faulkner and Runde [4], Kallinikos et al. [77], and Yoo et al. [32]. | • Adoption, diffusion, and infusion of digital technologies have led to some unexpected outcomes.<br>• Digital technologies have wrought both positive and negative outcomes that need to be better understood. |

**Note**: The column entries are often taken directly from the referenced document. In the interests of simplicity, however, we have omitted page numbers and often omitted quotation marks from most of the text in the table.

## APPENDIX B

### *Definitions of constructs in our theory*

Below we provide the definitions of constructs in our realist theory of digital objects, digital systems, and digitalized systems. The definitions are based on foundational ontological constructs that we summarize in Table 1 of our paper. Except for Definition 1 (Material Object), which we took directly from Bunge [[47], p. 22], the constructs below are the result of our adaptation of Bunge's theory of semantics and semiotics (e.g., [36,37,40]) and the extension of this theory to the context of digital phenomena.

| **Constructs of our realist theory of digital objects, digital systems, and digitalized systems** |
|---|
| *Definition 1:* ***Material Object*** *([47], p. 22)* |
| "An object $x$ is a *material object* (or *entity*) if, and only if, for every reference frame $y$, if $S_y(x)$ is a state space for $x$, then $S_y(x)$ contains at least two elements. Otherwise $x$ is an *immaterial object* (or *nonentity*)." |
| *Definition 2:* ***Sign*** |
| A *sign* is a thing that possesses a particular property (or properties) that maps to one or more other properties of the same thing, another object (thing or construct), or the properties of another object. |
| *Definition 3:* ***Symbol*** |
| A *symbol* is a thing that is an artificial, non-iconic sign. |
| *Definition 4:* ***Translator*** |
| A *translator* is a *thing* that maps (according to a set of rules) one object to another object. |
| *Definition 5:* ***Bitstring Construct*** |
| A *bitstring construct* is a mathematical *sequence* of zeros and ones (binary digits). |
| *Definition 6:* ***Instance of a Bitstring Construct*** |
| An instance of a *Bitstring Construct* is a specific sequence of zeros and ones. |
| *Definition 7:* ***Bitstring Symbol*** |
| A *bitstring symbol* is a thing that has at least one intrinsic, hereditary, or emergent property that maps the symbol to an instance of a bitstring construct. |
| *Definition 8a:* ***Digital Material Object (DMO)*** |
| A *digital material object* is an atomic digital symbol. |
| *Definition 8b:* ***Digital Conceptual Object (DCO)*** |
| A *digital conceptual object* is an atomic digital construct. |

(*continued on next page*)

(*continued*)

*Definition 9a:* ***Digital Material System (DMS)***
A *digital material system* is an artificial material system whose components are only digital material objects or other digital material systems.
*Definition 9b:* ***Digital Conceptual System (DCS)***
A *digital conceptual system* is an artificial conceptual system whose components are only digital conceptual objects or other digital conceptual systems.
*Definition 10a:* ***Digitalized Material System (DZMS)***
A *digitalized material system* is an artificial material system that has at least one component that is a digital material object, a digital material system, or another digitalized material system.
*Definition 10b:* ***Digitalized Conceptual System (DZCS)***
A *digitalized conceptual system* is an artificial conceptual system that has at least one component that is a digital conceptual object, a digital conceptual system, or another digitalized conceptual system.
*Definition 11:* ***Digital System, Digitalized System—Hereditary Property***
A *hereditary property* of a digital or digitalized system is a property in general that is also possessed by at least one of the system's components.
*Definition 12:* ***Digital System, Digitalized System—Emergent Property***
An *emergent property* of a digital or digitalized system is a property in general that is possessed only by the system and not any of its components.
*Definition 13:* ***Digital Object, Digital System, Digitalized System—Function***
A *function* of a digital object, digital system, and digitalized system is an outcome (a specific state of the digital object, digital system, and digitalized system) that it is designed to be achieved only via its *internal* processes.
*Definition 14:* ***Digital Object, Digital System, Digitalized System—Form***
The *form* of a digital object, digital system, or digitalized system is the set of all the properties that it possesses.
*Definition 15:* ***Digital Object, Digital System, Digitalized System—Mechanism***
The *mechanisms* of a digital object, digital system, or digitalized system are the internal processes that occur in them that enable them to enact their function.

## Data availability

No data was used for the research described in the article.

Roman Lukyanenko is an associate professor at the McIntire School of Commerce, University of Virginia. Roman’s research interests include conceptual modeling, information quality, crowdsourcing, machine learning, design science research, and research methods. Roman’s work on these topics has been published in Nature, MIS Quarterly, Information Systems Research, Journal of the Association for Information Systems, European Journal of Information Systems, among others.

Ron Weber is an emeritus professor at Monash University and The University of Queensland. His research interests lie in the areas of representation theory, conceptual modeling, ontology, and the philosophy of information systems.